\documentclass[11pt]{article}

\usepackage{amsfonts}
\usepackage{amsmath, amssymb, amscd}
\usepackage{amsthm}
\usepackage{graphicx}
\usepackage[mathscr]{eucal}
\usepackage{geometry}

\numberwithin{equation}{section}

\newtheorem{theorem}{Theorem}[section]

\theoremstyle{remark}
\newtheorem{remark}{Remark}[section]

\newcommand{\bfi}{\bfseries\itshape}
\DeclareMathOperator{\ext}{ext}

\usepackage{hyperref}
\usepackage[authoryear,sort&compress]{natbib}
\bibpunct{[}{]}{,}{n}{,}{,}

\begin{document}
\title{Generating Functionals and Lagrangian PDEs}
\author{Joris Vankerschaver\thanks{Current Address: Imperial College London, London SW7 2AZ, UK. Email: \url{Joris.Vankerschaver@gmail.com}.}, %
Cuicui Liao\thanks{Current Address: Department of Mathematics, Jiangnan University. No.1800 Lihu Avenue, Wuxi, Jiangsu, 214122, China. Email: \url{liaocuicuilcc@gmail.com}.}, %
Melvin Leok\thanks{Email: \url{mleok@math.ucsd.edu}.} \\[.5ex]
{\small Department of Mathematics} \\
{\small University of California, San Diego} \\
{\small 9500 Gilman Drive, Dept. 0112} \\
{\small La Jolla, CA 92093-0112, USA}
}
\maketitle

\begin{center}
{\it Dedicated to the memory of Jerrold E. Marsden.}
\end{center}

\begin{abstract}
	The main goal of this paper is to derive an alternative characterization of the multisymplectic form formula for classical field theories using the geometry of the space of boundary values.   We review the concept of Type-I/II generating functionals defined on the space of boundary data of a Lagrangian field theory.  On the Lagrangian side, we define an analogue of Jacobi's solution to the Hamilton--Jacobi equation for field theories, and we show that by taking variational derivatives of this functional, we obtain an isotropic submanifold of the space of Cauchy data, described by the so-called multisymplectic form formula.  As an example of the latter, we show that Lorentz's reciprocity principle in electromagnetism is a particular instance of the multisymplectic form formula. We also define a Hamiltonian analogue of Jacobi's solution, and we show that this functional is a Type-II generating functional.  We finish the paper by defining a similar framework of generating functions for discrete field theories, and we show that for the linear wave equation, we recover the multisymplectic conservation law of Bridges.
\end{abstract}

\section{Introduction}

In this paper, we revisit the ``multisymplectic form formula'' from \cite{MaPaSh1998}.  We show that this formula can be derived from infinite-dimensional symplectic geometry, and expresses the fact that the space of admissible boundary data for a given classical field theory is isotropic in the space of all Cauchy data, equipped with a certain symplectic form.  In the case of discrete field theories, defined on a discrete mesh in space-time, we show that these results have a natural discrete counterpart.  In particular, we derive the discrete multisymplectic form formula by considering the space of admissible boundary data within the space of (discrete) Cauchy data.

\paragraph{Geometry of the Space of Boundary Data.}

To  derive the expression for the boundary Lagrangian, we rely heavily on the results of Lawruk, \'Sniatycki and Tulczyjew \cite{LaSnTu1975}, who describe the space of boundary data associated to a given PDE,  showing among other things that a PDE determines an isotropic or sometimes even a Lagrangian submanifold of the space of all boundary data.  Kijowski and Tulczyjew \cite{KiTu1979} connect these results with the polysymplectic description of field theories, under the name of the ``finite domain description of field theories,''  and give a description in terms of \emph{generating forms} of Lagrangian submanifolds (see also \cite{SnTu1973}).   In the first two sections of this paper, we recall their results, described in the language of multisymplectic geometry.

In recent years, Rovelli \cite{Ro2004} has rediscovered many of these results in his search for a Hamiltonian description of field theories in which no space-time split is made.  He applied this formalism to general relativity using Ashtekar variables, rederiving among other things the Einstein--Hamilton--Jacobi equation.  Our approach agrees with Rovelli's results wherever appropriate, but one of the advantages of the multisymplectic approach used here is that we are easily able to tie in our results with infinite-dimensional symplectic geometry. 

In section~\ref{sec:boundary_lagrangian}, we recall from \cite{KiTu1979} that the action density of a classical field theory defines a generating function on the space of boundary data.  While the technical details are given below, the idea is that, for any given Lagrangian PDE and for any subset $U$ of the space of independent variables, we can define a functional $L_{\partial U}$, and by taking the exterior derivative, a one-form  
\begin{equation} \label{intro:dl}
	\mathbf{d} L_{\partial U}:  \mathscr{Y}_{\partial U}   \to    T^\ast \mathscr{Y}_{\partial U}
\end{equation}
which takes Dirichlet boundary data on $\partial U$ (the elements of $\mathscr{Y}_{\partial U}$) into the corresponding Neumann data along $\partial U$ (the elements of $T^\ast \mathscr{Y}_{\partial U}$).  
The image of this map turns out to be an isotropic submanifold of $T^\ast \mathscr{Y}_{\partial U}$ (and, depending on the geometry of the domain $U$ and the well-posedness of the underlying PDE, in some cases even a Lagrangian submanifold), so that the pull-back along $\mathbf{d} L_{\partial U}$ of the canonical symplectic form $\omega$ on $T^\ast \mathscr{Y}_{\partial U}$ vanishes:
\begin{equation}  \label{intro:isotropy}
	(\mathbf{d} L_{\partial U})^\ast \omega = 0.
\end{equation} 
Our main result in this section is then given in \eqref{msff}: by expressing that $\mathbf{d}^2 L_{\partial U}$ vanishes identically, we derive the so-called \emph{multisymplectic form formula} (see \cite{MaPaSh1998}), which we now discuss.

\paragraph{Generating Functionals and Multisymplecticity.}  

The multisymplectic form formula was introduced in \cite{MaPaSh1998} as a criterion for when the space of solutions of a PDE conserves a given multisymplectic form.  Concretely, in the case of field theories derived from a Lagrangian function $L$, the multisymplectic form formula takes on the following form.  If $\Omega_L$ is the \emph{Lagrangian multisymplectic form} (defined below in \eqref{PCform1} and \eqref{msform}), then for any subset $U$ of the space of independent variables, and for any solution $\phi$ of the Euler--Lagrange equations defined on $U$, we have that 
\begin{equation} \label{intro:MSFF}
	\int_{\partial U} (j^1 \phi)^\ast \left( \mathbf{i}_{j^1 W} \mathbf{i}_{j^1 V} \Omega_L \right) = 0,
\end{equation}
where $V$ and $W$ are arbitrary first variations of $\phi$.  If we restrict to vertical first variations, this formula takes on the following form in coordinates: 
\[
	\int_{\partial U} \frac{\partial^2 L}{\partial y^a_\mu \partial y^b_\nu} (V^a(x) W^b_{, \nu}(x) - W^a(x) V^b_{,\nu}(x)) d^n x_\mu = 0,
\]
where $V^a$, $W^b$ are the components of $V$ and $W$, and the subscript $\nu$ denotes differentiation with respect to $x^\nu$.

We note that Bridges \cite{Br1997} defined a different notion of multisymplecticity for Hamiltonian field theories.  In this context, a set of Hamiltonian PDEs is said to be multisymplectic if the equations satisfy a certain differential conservation law.  Under some restrictions, this conservation law can be rewritten in integral form to yield a result similar to \eqref{intro:MSFF}, but the precise link between both formulations in full generality is not yet clear.

As mentioned before, in this paper we take a different approach to the derivation of \eqref{intro:MSFF} and its implications.  Given a Lagrangian field theory, we construct the associated generating functional \eqref{intro:dl} and we relate this functional to the integral of the Poincar\'e--Cartan form $\Theta_L$ over the boundary $\partial U$.  In this way, we then show that the left-hand side of the isotropy condition \eqref{intro:isotropy}, when written out in terms of the multisymplectic form $\Omega_L = - \mathbf{d} \Theta_L$, is nothing but the multisymplectic form formula \eqref{intro:MSFF}.  

The advantage of this approach is that it provides a criterion of multisymplecticity which can be applied to any set of PDEs for which a Dirichlet-to-Neumann map can be defined.  From our point of view, therefore, the isotropy condition \eqref{intro:isotropy} is fundamental, and the multisymplectic form formula \eqref{intro:MSFF} arises as a consequence.

\paragraph{Generating Functionals.} 

A second aim of this paper is to review the theory of generating functionals for field theories (after \cite{SnTu1973} and \cite{KiTu1979}), in a way which allows for a straightforward generalization to discrete field theories.  In section~\ref{sec:genfunctional}, we show first that the boundary Lagrangian can be viewed as the field-theoretic analogue of a type-I generating function.   Our main result is then given in theorem~\ref{thm:msff}, where we show that the submanifold of admissible boundary conditions generated by the boundary Lagrangian is isotropic if and only if the so-called \emph{multisymplectic form formula} (see \cite{MaPaSh1998}) holds. After a few examples at the end of Section~\ref{sec:boundary_lagrangian} for which the boundary Lagrangian can be computed explicitly, we give a nontrivial example at the end of Section~\ref{sec:genfunctional} in the context of electromagnetism, where we show that Lorentz's principle of reciprocity can be seen as a particular example of the multisymplectic form formula.

We then set up the Hamiltonian formalism for field theories (following \cite{BiSnFi1988, CaCrIb1991}) and we show that the Hamiltonian function gives rise to a generating functional of type-II on the space of boundary data.  Both functionals, incidentally, can be viewed as field-theoretic analogues of Jacobi's solution to the Hamilton--Jacobi equation in mechanics.

This has important implications for discrete mechanics, since Jacobi's solution to the Hamilton--Jacobi equation is the exact discrete Lagrangian~\cite{MaWe2001}, and the order of approximation of a variational integrator can be characterized by the extent to which a computable discrete Lagrangian approximates the exact discrete Lagrangian. As such, a precise characterization of the boundary Lagrangian is essential for the construction and analysis of variational integrators for Lagrangian field theories.

\paragraph{Outlook: Discrete Lagrangian Field Theories.}

The boundary Lagrangian plays an analogous role in discrete Lagrangian field theories to the exact discrete Lagrangian of discrete variational mechanics. In particular, it was shown in \cite{LeSh2012} that the characterization of the exact discrete Lagrangian naturally leads to systematic methods for constructing computable discrete Lagrangians. Furthermore, the order analysis of variational integrators is significantly simplified by variational error analysis, which relates the order of a variational integrator with the order to which the discrete Lagrangian approximates the exact discrete Lagrangian. The development of a corresponding theory of variational error analysis for discretizations of Lagrangian field theories will rely, in part, on a deeper understanding of boundary Lagrangians, and how they serve as generating functionals for multisymplectic relations.

Another long-term goal is to clarify the concept of multisymplecticity for discrete Lagrangian field theories, in which the space of independent variables is replaced by a discrete mesh.  While both the criterion of Bridges (see \cite{BrRe2001, BrRe2006} and the references therein) as well as the multisymplectic form formula\footnote{It is interesting to note from a historical point of view that a precursor of the multisymplectic form formula already appears in the seminal work of Courant, Friedrichs and Lewy \cite{CoFrLe1928}.} can be defined in this context, the relation between both is not yet clear.  In the discrete context, we show that a generating functional akin to the one described in \eqref{intro:dl} can be introduced  and that its image determines an isotropic submanifold of the space of (finite-dimensional) discrete boundary data.  As in the case of continuous field theories, the condition of isotropy then gives rise to the discrete multisymplectic form formula.  While we have not yet been able to establish the link between this condition and Bridges' discrete version of multisymplecticity in full generality, we finish the paper with a simple discretization of the wave equation where this link can be established by direct computation.

\paragraph{Dedication.}
We dedicate this paper to the memory of Jerrold E. Marsden.  The methods pioneered by Jerry and his collaborators exerted a very profound influence on this paper:  for the treatment of the infinite-dimensional geometry of the space of boundary data, we relied heavily on the foundational results from \cite{AbMa1978, ChMa1974}, while the connection with field theory, and in particular the variational/multisymplectic formulation, uses the results from the GIMMSY manuscripts \cite{GIMMSY1,GIMMSY2}.  In more recent years, Jerry was influential in the development of variational principles for discrete field theories, his paper \cite{MaPaSh1998} being the first to give a definition of multisymplecticity for a discrete Lagrangian field theory.

\subsection*{The Geometry of Lagrangian Field Theories}  

In this section, we briefly recall the fiber bundle approach to classical field theory.  We give a description in local coordinates; for a intrinsic description, as well as applications and a more in-depth discussion, see \cite{BiSnFi1988, CaCrIb1991, CaIbLe1999, GIMMSY1, GIMMSY2} and the references therein.

Throughout this paper, we will consider fields as sections of a fiber bundle $\rho: Y \to X$.  Often, $X$ will be spacetime and $Y$ will be the product of $X$ with a vector space $V$, but this will not always be the case.  We will take $X$ to be orientable, of dimension $n + 1$ with $n \ge 0$, and we denote coordinates on $X$ by $(x^\mu)$, $\mu = 0, \ldots, n$.  We use the shorthand $d^{n + 1} x = dx^0 \wedge \cdots \wedge dx^n$, and we put 
\begin{align}
	d^n x_\mu  & := \mathbf{i}_{\partial/\partial x^\mu} d^{n+1}x \nonumber \\
		& \phantom{:} =  (-1)^\mu dx^0 \wedge \cdots \wedge dx^{\mu-1} \wedge dx^{\mu+1} \wedge \cdots \wedge dx^n,
			\label{dnxmu}
\end{align}
so that (up to sign) $d^n x_\mu$ is $d^{n+1} x$ with the $dx^\mu$ term removed.

On $Y$ we choose coordinates $(x^\mu, y^a)$, $a = 1, \ldots, k$, adapted to the projection $\rho$, so that a section $\phi: X \to Y$ can be locally written as $\phi(x) = (x^\mu, \phi^a(x))$, with $\phi^a(x)$ locally defined component functions. 

We define the first jet bundle $J^1 Y$ to consist of equivalence classes of local sections of $\pi$, where two sections $\phi, \phi': X \to Y$ are equivalent at a point $x \in X$ if their first-order Taylor expansions around $x$ agree, i.e., $T_x \phi = T_x \phi'$.  On $J^1 Y$, we have local coordinates $(x^\mu, y^a, y^a_\mu)$, where the $y^a_\mu$ can be considered as the derivatives of $y^a$ with respect to the variables $x^\mu$, which we refer to as multi-velocities.  Given a local section $\phi: X \to Y$ of $\rho$, we define the prolongation of $\phi$ to be the section $j^1 \phi$ of $J^1 Y$ given in local coordinates by 
\[
	j^1 \phi(x) = \left(x^\mu, \phi^a(x), \frac{\partial \phi^a}{\partial x^\mu}(x) \right),
\]
where the $\phi^a(x)$ are the component functions of $\phi$.

By a Lagrangian density on $J^1Y$ we mean a map $\mathcal{L} : J^1 Y \to \bigwedge^{n+1}(X)$, where $\bigwedge^{n+1}(X)$ is the space of volume forms on $X$.  Such a Lagrangian density can be written in local coordinates as $\mathcal{L}(x^\mu, y^a, y^a_\mu) = L(x^\mu, y^a, y^a_\mu) \, d^{n + 1} x$, where $L(x^\mu, y^a, y^a_\mu)$ is a function on $J^1 Y$ referred to as the Lagrangian function.  Throughout this paper, we will consider only first-order field theories, for which the Lagrangian density $\mathcal{L}$ depends on the first-order derivatives only.  For an overview of the issues that can occur for higher-order field theories, see \cite{Betounes1984}.  A variational description of higher-order field theories was given in \cite{KoSh2000}.

Given a Lagrangian density $\mathcal{L}$, we can introduce a number of geometric objects on the first jet bundle.  In local coordinates, the Poincar\'e--Cartan form is given by 
\begin{equation} \label{PCform1}
	\Theta_L  = \left( L  - \frac{\partial L}{\partial y^a_\mu} y^a_\mu \right)  d^{n + 1} x 
		+ \frac{\partial L}{\partial y^a_\mu} dy^a \wedge d^n x_\mu.
\end{equation}	
We define the multisymplectic form $\Omega_L$ on $J^1 Y$ by 
\begin{equation} \label{msform}
	\Omega_L = - \mathbf{d} \Theta_L.
\end{equation}

The expression \eqref{PCform1} for the Poincar\'e--Cartan form can be rewritten as 
\begin{equation} \label{PCform2}
	\Theta_L = L d^{n + 1} x + \frac{\partial L}{\partial y^a_\mu} (dy^a - y^a_\nu dx^\nu) \wedge d^n x_\mu,
\end{equation}
since from \eqref{dnxmu} it follows that $dx^\nu \wedge d^n x_\mu = \delta^\nu_\mu d^{n+1}x$, with $\delta^\nu_\mu$ the Kronecker delta.  The advantage of this expression is that the forms $dy^a - y^a_\nu dx^\nu$ which appear in the second term are \emph{contact forms}, i.e., they vanish when pulled back along a prolongated section: if $j^1 \phi(x) = (x, \phi^a(x), \partial \phi^a(x)/\partial x^\mu)$ in local coordinates, then
\[
	(j^1 \phi)^\ast (dy^a - y^a_\nu dx^\nu) = d \phi^a(x) - \frac{\partial \phi^a(x)}{\partial x^\nu} dx^\nu = 0.
\]
Along prolongated sections, we therefore have that 
\begin{equation} \label{L_relation}
	(j^1 \phi)^\ast \Theta_L = (j^1 \phi)^\ast (L d^{n+1}x).
\end{equation}
This equality will often be useful later on.

\section{The Space of Boundary Data}

Let $U$ be an open subset of $X$ with boundary $\partial U$.  We want to prescribe boundary data along $\partial U$ with values in $Y$, the total space of the configuration bundle $\rho: Y \to X$.  We first describe the geometry of the space of all boundary data, and then we discuss some related spaces.  We emphasize that at this stage, $\partial U$ does not have to be a Cauchy surface, or even be spacelike: indeed, all of the definitions below are independent of the choice of a metric on $X$, so that in particular they can be applied to hyperbolic and elliptic problems alike.  Most of the material in this section can be found in \cite{SnTu1973, LaSnTu1975}; for applications to more general field theories, see \cite{BiSnFi1988, KiTu1979}.

By an {\bfi element of boundary data} on $U$, we mean a section $\varphi: \partial U \to Y$ of $\rho$, defined on the boundary of $U$.  We denote by $\mathscr{Y}_{\partial U}$ the space of all boundary data and we now describe the tangent and cotangent bundles of this space.  We can describe the tangent vectors $\delta \varphi$ at a point   $\varphi \in \mathscr{Y}_{\partial U}$ as follows: let $\varphi_\epsilon$ be a curve in $\mathscr{Y}_{\partial U}$ such that $\varphi_{\epsilon = 0} = \varphi$, and put 
\[
	\delta \varphi(x) := \frac{d \varphi_\epsilon(x)}{d\epsilon} \Big|_{\epsilon = 0}
\]
for all $x \in \partial U$.  Note that the right-hand side takes values in the space of vertical tangent vectors: as each map $\varphi_\epsilon$ is a section of $\rho$, we have that $\rho \circ \varphi_\epsilon = \mathrm{Id}$,  and therefore
\[
	T \rho ( \delta \varphi(x) )  = \frac{d}{d \epsilon}\Big|_{\epsilon = 0} \rho(\varphi_\epsilon(x)) = 0.
\]

As the right-hand side takes values in $V_{\varphi(x)} Y$, we have that $\delta \varphi$ is a map from $\partial U$ into $VY$ with the property that  $\delta \varphi(x) \in V_{\varphi(x)} Y$ for all $x \in \partial U$.   These maps can alternatively be described as vector fields along $\varphi$, or as sections of the pullback bundle $\varphi^\ast(VY)$.  In any case, we have that the tangent space at a point is given by 
\[
	T_\varphi \mathscr{Y}_{\partial U} = \left\{ \delta \varphi: \partial U \to VY : \delta \varphi(x) \in V_{\varphi(x)} Y \quad \text{for all $x \in \partial U$} \right\}.
\]
In local coordinates, an element $\delta \varphi$ of $T_\varphi \mathscr{Y}_{\partial U}$
is given by 
\[
	\delta \varphi(x) = \delta \varphi^a(x) \frac{\partial}{\partial y^a},
\]
where the $\delta \varphi^a(x)$ are locally defined component functions.   Roughly speaking, we can think of the elements $\delta \varphi \in T_\varphi \mathscr{Y}_{\partial U}$ as \emph{infinitesimal variations} of the boundary data given by $\varphi$.  Following \cite{BiSnFi1988}, we will refer to $T\mathscr{Y}_{\partial U}$ as the space of {\bfi Cauchy data}.

We now describe the dual spaces $T^\ast_\varphi \mathscr{Y}_{\partial U}$.  We restrict ourselves to the smooth dual, in other words, the space of smooth linear functionals from $T_\varphi \mathscr{Y}_{\partial U}$ to $\mathbb{R}$.  Since the elements of $T_\varphi \mathscr{Y}_{\partial U}$ are vertical vector fields along $\partial U$, the elements of the dual can be identified with linear maps $\pi$ from the space of vertical vector fields to the space of densities on the boundary, so that the duality pairing can be defined in terms of integration over the boundary by 
\begin{equation} \label{duality_pairing}
	\left< \delta {\varphi}, \pi \right> = \int_{\partial U} \pi \cdot \delta {\varphi}.
\end{equation}
Here, $\pi \cdot \delta \varphi$ is the volume form on $\partial U$ obtained by letting the linear map $\pi$ act on the vertical vector field $\delta \varphi$.

We now make this picture more precise.  To describe $T^\ast_\varphi \mathscr{Y}_{\partial U}$, we consider first the tensor product bundle $\varphi^\ast(V^\ast Y) \otimes \bigwedge^{n}(\partial U)$, where $V^\ast Y$ is the dual bundle to the vertical bundle $VY$,  $\varphi^\ast(V^\ast Y)$ is the pullback bundle over $\partial U$, and $\bigwedge^{n}(\partial U)$ is the bundle of volume forms (densities) on the boundary $\partial U$, viewed as a bundle over $Y$.    Note that if there is a locally defined volume form $dS$ on $\partial U$, then the elements of $\varphi^\ast(V^\ast Y) \otimes \bigwedge^{n}(\partial U)$ can be written locally as $\pi = \pi_a dy^a \otimes dS$.  

We now let $T^\ast_\varphi \mathscr{Y}_{\partial U}$ be the space of sections of $V^\ast Y \otimes \bigwedge^{n}(\partial U)$, defined on the boundary $\partial U$.  Each section $\pi \in T^\ast_\varphi \mathscr{Y}_{\partial U}$ is a map from $\partial U$ into $\varphi^\ast(V^\ast Y) \otimes \bigwedge^{n}(\partial U)$, with the property that 
\[
	\pi(x) \in V^\ast_{\varphi(x)} Y \otimes  \mbox{$\bigwedge^{n}_x$}(\partial U)
\]
for all $x \in \partial U$.  Again under the assumption that $dS$ is a local volume form on $\partial U$, we have that  $\pi$  can be written locally as 
\begin{equation} \label{instantmom}
	\pi(x) = \pi_a dy^a \otimes dS.
\end{equation}
For reasons that will become clear later on, we will refer to $\pi$ as the momentum in the direction normal to $\partial U$, and we will refer to the cotangent bundle $T^\ast \mathscr{Y}_{\partial U}$ as the space of {\bfi normal momenta}.

The duality pairing \eqref{duality_pairing} between $T \mathscr{Y}_{\partial U}$ and $T^\ast \mathscr{Y}_{\partial U}$ can then be written as 
\[
	\left< \delta {\varphi}, \pi \right> = \int_{\partial U} \pi \cdot \delta {\varphi} 
	 = \int_{\partial U} \pi_a(x) \delta \varphi^a(x) \, dS.
\]

For future reference, we point out that $T^\ast \mathscr{Y}_{\partial U}$ is equipped with a canonical weak symplectic form, given by $\omega = - \mathbf{d} \Theta$, where $\Theta$ is the one-form defined intrinsically as 
\begin{equation} \label{canonform}
	\Theta(\pi)(\delta \pi) = \left< T \tau( \delta \pi), \pi \right>,
\end{equation}
for all $\pi \in T^\ast \mathscr{Y}_{\partial U}$ and $\delta \pi \in T_\pi (T^\ast \mathscr{Y}_{\partial U})$,   where $\tau:T^\ast \mathscr{Y}_{\partial U} \to  \mathscr{Y}_{\partial U}$ is the cotangent bundle projection (see \cite{AbMa1978}).  Instead of this definition, we will often use the following defining property of $\Theta$: for every one-form $\beta$ on $\mathscr{Y}_{\partial U}$, we have that 
\begin{equation} \label{pullback_property}
	\beta^\ast \Theta = \beta,
\end{equation}
where on the left-hand side $\beta$ is interpreted as a map from $\mathscr{Y}_{\partial U}$ into $T^\ast \mathscr{Y}_{\partial U}$.

\section{The Boundary Lagrangian} \label{sec:boundary_lagrangian}

Let $\mathcal{L}: J^1 Y \to \bigwedge^{n+1}(X)$ be a first-order Lagrangian density and denote, as before, $\mathcal{L}(j^1 \phi) = L(j^1 \phi) \, dV$, with $L(j^1 \phi)$ a scalar function on $J^1 Y$.  To make the distinction between boundary data on $\partial U$ and fields defined on the interior of $U$, we will denote the former by $\varphi \in \mathscr{Y}_{\partial U}$, while the latter are denoted by $\phi$.

We now define, after \cite{KiTu1979},  the {\bfi boundary Lagrangian} $L_{\partial U}$ as the functional on $\mathscr{Y}_{\partial U}$ given by 
\begin{equation} \label{boundarylagrangian}
	L_{\partial U}(\varphi)=\int_{U}L(j^{1}\phi)dV,
\end{equation}
where $\phi$ is a (not necessarily unique) section satisfying the Euler--Lagrange equations, and such that $\phi$ agrees with $\varphi$ on $\partial U$: 
\begin{equation} \label{ELeqnswithboundarydata}
	\frac{d}{dx^\mu} \left( \frac{\partial L}{\partial y^a_\mu}(j^1 \phi) \right) - 
		\frac{\partial L}{\partial y^a}(j^1 \phi) = 0
	\quad \text{and} \quad \phi_{|\partial U} = \varphi.
\end{equation}
In other words, 
\begin{equation} \label{Laction}
	L_{\partial U}(\varphi) = \mathcal{S}(\phi),
\end{equation}
where $\mathcal{S}$ is the action functional, and $\varphi$ and $\phi$ are related as described in \eqref{ELeqnswithboundarydata}.  This leads us to an alternative description of the boundary Lagrangian, which will often be useful in computations: as the solutions $\phi$ of the Euler--Lagrange equations are precisely the extremal points of the action functional $\mathcal{S}$, we see that $L_{\partial U}(\varphi)$ is precisely the extremal value of $\mathcal{S}$:
\begin{equation} \label{Lext}
	L_{\partial U}(\varphi) = \underset{\phi_{|\partial U} = \varphi}{\ext} \mathcal{S}(\phi),
\end{equation}
where the extremal value of $\mathcal{S}$ is computed over the class of all sections $\phi$ such that $\phi_{|\partial U} = \varphi$. 

\begin{remark} \label{rem:degenerate_lagrangian}
In general, the Euler--Lagrange boundary-value problem might not uniquely determine a section of the configuration bundle. If one considers the critical surface of sections satisfying the Euler--Lagrange boundary-value problem, then every section in a connected component of the critical surface yields the same action. As such, the issue of non-uniqueness only poses an issue in the definition of the boundary Lagrangian if there is more than one connected component to the critical surface, and this affects both of the characterizations above equally. We can ensure that there is only one connected component to the critical surface by requiring that the domain $U$ in the definition \eqref{boundarylagrangian} of $L_{\partial U}(\varphi)$ is sufficiently small. A similar constraint also ensures that the Euler--Lagrange equations \eqref{ELeqnswithboundarydata} have a solution on $U$.
\end{remark}

\paragraph{The Space of Admissible Boundary Data.}  
It will often happen that one cannot prescribe arbitrary boundary data for the Euler--Lagrange equations \eqref{ELeqnswithboundarydata}.  We will denote by $\mathscr{K}_{\partial U}$ the space of boundary data that give rise to a unique solution of the Euler--Lagrange equations, and we refer to $\mathscr{K}_{\partial U} \subset \mathscr{Y}_{\partial U}$ as the space of {\bfi admissible boundary data}.  The specification of $\mathscr{K}_{\partial U}$ will depend on the type of PDE under study and the geometry of $U$.  We give some examples at the end of this section.

\paragraph{Functional Derivatives.}
We define the functional derivative of $L_{\partial U}$ as follows: $\delta L_{\partial U}/\delta \varphi$ is the unique element of $T^\ast \mathscr{Y}_{\partial U}$ such that 
\[
	\mathbf{d} L_{\partial U}(\varphi) \cdot \delta \varphi = 
		\int_{\partial U} \frac{\delta L_{\partial U}}{\delta \varphi} \cdot \delta \varphi
\]
for every variation $\delta \varphi \in T \mathscr{Y}_{\partial U}$, where $\mathbf{d} L_{\partial U}$ is the exterior derivative of the boundary Lagrangian.  If not all boundary data are admissible, so that $\mathscr{K}_{\partial U}$ is a proper subset of $\mathscr{Y}_{\partial U}$, then the admissible variations $\delta \varphi$ are elements of $T \mathscr{K}_{\partial U}$, so that $\mathbf{d} L_{\partial U}(\varphi) \in T_\varphi^\ast \mathscr{K}_{\partial U}$.

By applying the exterior differential to both sides of \eqref{Laction}, we now obtain 
\begin{equation} \label{relation_derivatives}
	\mathbf{d} L_{\partial U}(\varphi) \cdot \delta \varphi
	= \mathbf{d} \mathcal{S}(\phi) \cdot \delta \phi,
\end{equation}
where $\phi$ is the solution of the Euler--Lagrange equations with boundary data $\varphi$, and $\delta \phi$ is a {\bfi first variation} of $\phi$, defined as follows.  Let $\varphi_\epsilon$ be a curve in the space $\mathscr{Y}_{\partial U}$ of boundary data such that $\varphi_{\epsilon = 0} = \varphi$ and $\frac{d}{d\epsilon} \varphi_\epsilon \big|_{\epsilon = 0} = \delta \varphi$, and let $\phi_\epsilon$ be the corresponding family of solutions of the Euler--Lagrange equations such that $(\phi_\epsilon)_{|\partial U} = \varphi_\epsilon$.  Then $\delta \phi$ is given by 
\begin{equation} \label{eq_firstvar}
	\delta \phi(x) = \frac{d\phi_\epsilon(x) }{d\epsilon}  \Big|_{\epsilon = 0},
\end{equation}
for all $x \in \partial U$.
Explicitly, the first variations $\delta \phi$ satisfy the first-variation equation, obtained by linearizing the Euler--Lagrange equations.

The exterior differential of the action functional is given by the first variation formula:
\begin{align*}
	\mathbf{d} \mathcal{S}(\phi) \cdot \delta \phi & = 
		\int_U \left( 
			\frac{\partial L}{\partial y^a} \delta y^a + \frac{\partial L}{\partial y^a_\mu} \delta y^a_\mu 
			\right) \, d^{n+1} x \\
		& = \int_U \left( 
			\frac{\partial L}{\partial y^a}  - \frac{d}{dx^\mu} \left(\frac{\partial L}{\partial y^a_\mu}  
			\right) \right) \delta y^a \, d^{n+1} x
			+ \int_{\partial U} \frac{\partial L}{\partial y^a_\mu} \delta y^a \, d^nx_\mu. 
\end{align*}
As $\phi$ is a solution of the Euler--Lagrange equations, the integral over $U$ vanishes and we conclude that 
\begin{equation} \label{derformula}
	\mathbf{d} L_{\partial U}(\varphi) \cdot \delta \varphi
	=
	\int_{\partial U} \frac{\partial L}{\partial y^a_\mu} \delta y^a \, d^nx_\mu,
\end{equation}
and therefore 
\begin{equation} \label{funcder}
	\frac{\delta L_{\partial U}}{\delta \varphi} = 
	   \frac{\partial L}{\partial y^a_\mu} dy^a \otimes \iota^\ast \left( d^nx_\mu \right), 
\end{equation}
where $\iota : \partial U \hookrightarrow X$ is the embedding of the boundary $\partial U$ into $X$.  Since $d^n x_\mu$ is an $n$-form on $X$, its pullback along $\iota$ is a form of maximal degree on $\partial U$.  If we choose coordinates on $X$ that are adapted to $\partial U$, in the sense that $\partial U$ is locally given by $x^0 = 0$, then $\iota$ is locally given by $\iota(x^1, \ldots, x^n) = (0, x^1, \ldots, x^n)$, so that 
\[
\iota^\ast \left( d^nx_0 \right) =  d^n x_0, \quad \text{and} \quad 
\iota^\ast \left( d^nx_i \right) = 0 \quad (i = 1, \ldots, n).
\]
As a result, in adapted coordinates we have that 
\[
\frac{\delta L_{\partial U}}{\delta \varphi} = \frac{\partial L}{\partial y^a_0} dy^a \otimes d^nx_0.
\]
A more intrinsic expression may be given when $X$ is equipped with a metric tensor, as we now show.

\paragraph{Normal Momenta.}  
When a Riemannian or Lorentzian metric $G$ on $X$ is given, we may describe the functional derivatives as follows.  In both cases, we have that 
\[
	\iota^\ast (d^n x_\mu) = n_\mu \, dS,
\]
(see \cite{Wa1984}), where $n^\mu$ is the outward normal to $\partial U$, $d S$ is the induced metric volume form on $\partial U$, and indices are raised/lowered using the metric.  The functional derivatives \eqref{funcder} can then be written as 
\begin{equation}\label{normalmomentum}
	\pi := \frac{\delta L_{\partial U}}{\delta \varphi}
	= \frac{\partial L}{\partial y^a_\mu} n_\mu \, dy^a \otimes dS,
\end{equation}
so that by comparing with \eqref{instantmom}, we have for the components 
\begin{equation} \label{components_normal_momentum}
	\pi_a =  \frac{\partial L}{\partial y^a_\mu} n_\mu = p_a^\mu n_\mu.
\end{equation} 
In other words, the boundary momentum $\pi_a$ is the normal component of the spacetime momentum $p_a^\mu$, so that we will refer to $\pi \in T^\ast \mathscr{Y}_{\partial U}$ as the {\bfi normal momentum} to the boundary $\partial U$.

\paragraph{Multisymplectic Form Formula.}  We now arrive at the first main result of this paper: a symplectic derivation of the multisymplectic form formula \eqref{intro:MSFF} using the geometry of the space of boundary data.  We first need a more careful derivation of \eqref{derformula}.  The boundary Lagrangian $L_{\partial U}$ can be written in terms of the Poincar\'e--Cartan form $\Theta_L$ as 
\[
		L_{\partial U}(\varphi) = \int_{\partial U} (j^1 \phi)^\ast \Theta_L
\]
where $\phi$ is the solution of the Euler--Lagrange equations with boundary data $\varphi$, and $j^1 \phi$ is its first jet prolongation.  Here, we have used \eqref{L_relation} to bring in the Poincar\'e--Cartan form.

By taking the exterior derivative, we then obtain 
\begin{equation} \label{dl}
\mathbf{d} L_{\partial U}(\varphi) \cdot \delta \varphi
	=
	\int_{\partial U} (j^1 \phi)^\ast \left( \mathbf{i}_{j^1 V} \Theta_L \right),
\end{equation}
where the vector field $V$ is a first variation of the solution $\phi$, defined as before in \eqref{eq_firstvar}. The advantage of this expression is that we can now take the exterior derivative again, to obtain 
\begin{equation} \label{ddl}
	\mathbf{d}^2 L_{\partial U}(\varphi) \cdot (\delta \varphi, \delta \varphi') = 
		\int_{\partial U} (j^1 \phi)^\ast \left( \mathbf{i}_{j^1 W} \mathbf{i}_{j^1 V} \Omega_L \right), 
\end{equation}
where $V$ and $W$ are the first variations induced by $\delta \varphi$ and $\delta \varphi'$, respectively.  The proof of this results proceeds along similar lines as the proof of Lemma~5.1 in  \cite{GIMMSY2}.  Since $\mathbf{d}^2 \equiv 0$, we now have that 
\begin{equation} \label{msff}
	\int_{\partial U} (j^1 \phi)^\ast \left( \mathbf{i}_{j^1 W} \mathbf{i}_{j^1 V} \Omega_L \right) = 0,
\end{equation}
for all solutions $\phi$ of the Euler--Lagrange equations, and all first variations $V, W$.  This is the {\bfi multisymplectic form formula}, first proposed in \cite{MaPaSh1998}.   Our derivation is close in spirit to the one in that paper, because of the link between the boundary Lagrangian and the action functional.  In Section~\ref{sec:genfunctional} we will see the multisymplectic form formula appear under a different guise, as the condition for the manifold of physical solutions to be an isotropic submanifold of the space of normal momenta.  

We finish by noting that our version of the multisymplectic form formula is somewhat less general than the one derived in \cite{MaPaSh1998}, since we consider only vertical variations. However, as one of the motivations for this work is the derivation of a multisymplectic form formula for discrete field theories, for which infinitesimal horizontal variations are not really well-defined, this is not a fundamental restriction.

\paragraph{Relation with the Crnkovi\'c--Witten Symplectic Form.}  As pointed out by Rovelli \cite{Ro2004}, the Crnkovi\'c--Witten symplectic form on the solution space can be related to the various structures on the space of boundary data.  We present here a slightly different approach from Rovelli, emphasizing the link with the multisymplectic form formula.  

We assume that the base space $X$ is equipped with a Lorentzian metric and we let $U$ be a region bounded by two spacelike hypersurfaces, $\Sigma_+$ and $\Sigma_-$.  We choose the orientation so that $\partial U = \Sigma_+ - \Sigma_-$, and we consider a boundary Lagrangian $L_{\partial U}$ for this particular geometry.  In Wheeler's terminology \cite{MiThWh1973}, $U$ would be called a ``thick sandwich.''

From the expression \eqref{iso_to_msff}, to be proved below, for the pullback $(\mathbf{d} L_{\partial U})^\ast \omega$, or alternatively from the multisymplectic form formula \eqref{msff} directly, we have that 
\[
\int_{\Sigma_-} (j^1 \phi)^\ast \left( \mathbf{i}_{j^1 W} \mathbf{i}_{j^1 V} \Omega_L \right)
=
\int_{\Sigma_+} (j^1 \phi)^\ast \left( \mathbf{i}_{j^1 W} \mathbf{i}_{j^1 V} \Omega_L \right),
\]
for all solutions $\phi$ of the Euler--Lagrange equations, and first variations $V, W$.  These expressions, however, are nothing but the symplectic structures of Crnkovi\'c--Witten \cite{CrWi1987} and Zuckerman \cite{Zu1987} on the space of solutions associated to $L$, integrated respectively over $\Sigma_-$ and $\Sigma_+$. We have therefore shown that the definition of this symplectic structure is independent of the spatial hypersurface along which to integrate.  This conclusion is of course not new, but the link with the multisymplectic form formula has hitherto not been established.

Finally, we also remark that Garc\'{\i}a \cite{Ga1974} has introduced a similar presymplectic form on the solution space of a given Lagrangian field theory.

\subsection*{Examples}  

For most field theories, the boundary Lagrangian  \eqref{boundarylagrangian} cannot be computed explicitly.  Here we present a number of examples where this is possible after all.  We stress, however, that the multisymplectic form formula and other theoretical results still remain valid (and yield useful information) even if we do not have an explicit expression for the boundary Lagrangian: at the end of Section~\ref{sec:genfunctional} below, we treat the example of electromagnetism and we show that the multisymplectic form formula is equivalent to the statement of Lorentz reciprocity.

\paragraph{Mechanics.}

In the case of a mechanical system on a configuration space $Q$ and with regular Lagrangian $L(q, v): TQ \to \mathbb{R}$, we let $X$ be $\mathbb{R}$ and we take for $Y$ the product $\mathbb{R} \times Q$.  The projection $\rho: Y \to X$ is then the projection onto the first factor.

We let $U$ be an open subset of $X = \mathbb{R}$ and we assume without loss of generality that $U$ is an interval $(0, h)$, so that $\partial U = \{ 0, h \}$.  In this case, the space of boundary data is just the product $Q \times Q$, where we think of the first, resp. the second factor, as specifying the configuration of the system at $t = 0,  h$, respectively.  Accordingly, the space of normal momenta is given by $T^\ast Q \times T^\ast Q$.  If $L$ is nondegenerate and $h$ is small enough, then it is well known that for every pair $(q_0, q_1)$ in $Q \times Q$ there exists a unique solution $q(t)$ of the Euler--Lagrange equations so that $q(0) = q_0$ and $q(h) = q_1$ (see, for instance, \cite{MaWe2001}).

According to the definition \eqref{boundarylagrangian}, the boundary Lagrangian is then given by 
\[
	L_{\partial U}(q_0, q_1) := \int_{0}^{h} L(q(t), \dot{q}(t)) dt,
\]
where $q(t)$ is the unique solution of the Euler--Lagrange equations with boundary data $(q_0, q_1)$.  In this way, we recover the \emph{exact discrete Lagrangian} introduced in \cite{MaWe2001}.  The variational derivatives \eqref{funcder} are  
\[
	\frac{\delta L_{\partial U}}{\delta q_0} = 
		- \frac{\partial L}{\partial \dot{q}}(q(0), \dot{q}(0)), \quad \text{and} \quad \frac{\delta L_{\partial U}}{\delta q_1} = 
		\frac{\partial L}{\partial \dot{q}}(q(h), \dot{q}(h)),
\]
and coincide with the momenta of the system at the begin and end point of the solution trajectory $q(t)$, $t \in [0, h]$.  The minus sign is due to the orientation of the boundary $\{0, h\}$. Notice that these equations are precisely the implicit characterization of a symplectic map, where $L_{\partial U}$ is viewed as a Type-I generating function.

\paragraph{Harmonic Functions.} 
Consider secondly the case of harmonic functions on $\mathbb{R}^n$, for which the Lagrangian is given by 
\[
	L(\phi, \nabla \phi) = \frac{1}{2} \left\Vert \nabla \phi \right\Vert^2 
\]
so that the field equations are given by Laplace's equation, $\Delta \phi = 0$.  The action is then given by 
\[
	\mathcal{S}(\phi) = \frac{1}{2} \int_U \left\Vert \nabla \phi \right\Vert^2 \, dV = \frac{1}{2} \int_{\partial U}  \phi \frac{\partial \phi}{\partial n} \, dl 
	- \frac{1}{2} \int_U \phi \Delta \phi \, dV,
\]
where we have used the divergence theorem, with $\partial \phi/\partial n := n \cdot \nabla \phi$ the normal derivative.  If we let $\phi$ be a solution of Laplace's equation with prescribed boundary data $\varphi$ on $\partial U$, we obtain for the boundary Lagrangian \eqref{boundarylagrangian}
\begin{equation} \label{harmonicBL}
	L_{\partial U}(\varphi) = 
		\frac{1}{2} \int_{\partial U}  \varphi \frac{\partial \phi}{\partial n} \, dl,
\end{equation}
so that the variational derivative becomes 
\[
	\frac{\delta L_{\partial U}}{\delta \varphi}
		= \frac{\partial \phi}{\partial n} \, d\phi \otimes dl,
\]
where $dl$ is the Euclidian line element along $\partial U$.
We conclude that the map which associates to each element of boundary data $\varphi$ the corresponding variational derivative $\delta L_{\partial U}/\delta \varphi$ is nothing but the \emph{Dirichlet-to-Neumann map} of the Laplace equation.      Last, if $n \ge 3$,  the boundary Lagrangian \eqref{boundarylagrangian} can also be formally related to the \emph{harmonic capacity} of the domain $U$.

\paragraph{The Wave Equation.} 

Our last example concerns the wave equation $\phi_{tt} - \phi_{xx} = 0$, with Lagrangian 
\begin{equation} \label{wave_lagrangian}
	L(\phi, \phi_{, \mu}) = \frac{1}{2} (\phi_{,t}^2 - \phi_{,x}^2 ).  
\end{equation}	
Our exposition follows \cite{KiTu1979, LaSnTu1975}.  We assume that $U$ is a square in $\mathbb{R}^2$ of unit length whose corner vertices are given by $(0, 0), (1, 0), (1, 1), (0, 1)$ and we prescribe boundary data $\varphi(t, x)$ along $\partial U$.  An arbitrary solution to the linear wave equation can be written as $\phi(t, x) = F(x-t) + G(x+t)$, where $F$ and $G$ are determined through the boundary conditions by 
\[
	F(x) + G(x) = \varphi(0, x), \quad F(x-1) + G(x+1) = \varphi(1, x), 
\]
and
\[
	F(-t) + G(t) = \varphi(t, 0), \quad F(1-t) + G(1+t) = \varphi(t, 1).
\]

Note that for $F$ and $G$ to be determined from this set of equations, the boundary data needs to satisfy the following compatibility condition:
\begin{equation}
	\varphi(t, 0) + \varphi(1-t, 1) - \varphi(0, x) - \varphi(1, 1-x) = 0.
\end{equation}
For the unit square $U$, the space of admissible boundary data $\mathscr{K}_{\partial U}$ is therefore the space of all smooth functions on the boundary $\partial U$ that satisfy this condition.  By substituting the solution of the wave equation with given boundary data $\varphi$ back into the action density \eqref{wave_lagrangian}, we then obtain the following expression for the boundary Lagrangian: 
\[
	L_{\partial U}(\varphi) = \int_0^1 \big( \varphi_x(\alpha, 0) - \varphi_t(0, \alpha) \big)
		\big(\varphi(1-\alpha, 1) - \varphi(0, \alpha) \big) \, d \alpha.
\]
If the boundary of $U$ consists (partly) of characteristic curves of the wave equation, then the wave equation will still be well-posed, but more restrictive compatibility conditions will arise.  Assume, for instance, that $U$ is the diamond shape in the $(t, x)$-plane bounded by the lines $x \pm t = 0$ and $x \pm t = 1$.  It will then only be possible to prescribe arbitrary boundary conditions along two adjacent sides of the diamond.  In \cite{KiTu1979}, it is shown that the image of $\mathbf{d} L_{\partial U}(\varphi)$ will in this case only be isotropic, while for the non-characteristic square the image of $\mathbf{d} L_{\partial U}(\varphi)$ can be shown to be Lagrangian.  We will not dwell on the continuous case any further and refer instead to \cite{KiTu1979, LaSnTu1975}, but later on we will see that a similar dichotomy arises in the discrete context.

\section{Type-I/II Generating Functionals}  \label{sec:genfunctional}

The purpose of this section is to show that the boundary Lagrangian $L_{\partial U}$ can be viewed as a generating functional, in some appropriate sense.  We proceed along the following route: first, we take variational derivatives to define a mapping from $\mathscr{Y}_{\partial U}$ to $T^\ast \mathscr{Y}_{\partial U}$, given by 
\[
	\mathbf{d} L_{\partial U}: \varphi \in \mathscr{Y}_{\partial U} 
		\mapsto 
		\frac{\delta L_{\partial U}}{\delta \varphi}(\varphi) \in T^\ast \mathscr{Y}_{\partial U}
\]
and generated by $L_{\partial U}$.   That is, $\mathbf{d} L_{\partial U}(\varphi)$ gives us the normal momenta $\delta L_{\partial U}/\delta \varphi$, given the fields $\varphi$ at the boundary.  Secondly, we show that the image of this map is a Lagrangian submanifold of $T^\ast \mathscr{Y}_{\partial U}$ with respect to the canonical symplectic structure \eqref{canonform}.  Since $\mathbf{d} L_{\partial U}$ is an exact one-form, this will be a straightforward consequence of the property \eqref{pullback_property}.  Based on these two properties, we will say that $L_{\partial U}$ is an example of a Type-I generating functional.  

In the second part of this section, we show that the image of $\mathbf{d} L_{\partial U}$ can be generated by other functionals as well.  More precisely, we can imagine subdividing the boundary $\partial U$ into two subsets $A$ and $B$, and prescribing boundary data $\varphi_A$ along $A$ and normal momenta $\pi_B$ along $B$.  We can then define a functional $H_{\partial U}(\varphi_A, \pi_B)$ of these data, whose variational derivative with respect to $\varphi_A$ is the normal momentum along $A$, and whose variational derivative with respect to $\pi_B$ is the field $\varphi$ along $B$. Consequently, the image of $\mathbf{d} \mathfrak{S}_{\mathrm{II}}$ coincides with the image of $\mathbf{d} H_{\partial U}$ in $T^\ast \mathscr{Y}_{\partial U}$, and in analogy with mechanics, we will call $H_{\partial U}$ a Type-II generating functional.  

\paragraph{Geometry of Generating Functionals.} 

Let $\mathfrak{S} : \mathscr{Y}_{\partial U} \to \mathbb{R}$ be an arbitrary functional on the space of boundary data.  The exterior derivative $\mathbf{d} \mathfrak{S}$ is a closed form, and its image is a Lagrangian submanifold of $T^\ast \mathcal{Y}_{\partial U}$.  If instead we restrict $\mathbf{d} \mathfrak{S}$ to a subspace $\mathscr{K}_{\partial U}$ of  $\mathscr{Y}_{\partial U}$, the resulting image is no longer Lagrangian, and we obtain merely an isotropic submanifold.     We collect these observations in the following theorem, the proof of which is a combination of standard results in symplectic geometry (see, for instance, \cite{AbMa1978}).   For the remainder of this section, we will tacitly assume that all infinite-dimensional spaces can be made into manifolds.  Under this assumption, the following standard theorem follows easily from the fact that $\mathbf{d}^2 \mathfrak{S} \equiv 0$.

\begin{theorem} \label{thm:isotropy}
	Let $\mathscr{K}_{\partial U} \subset  \mathscr{Y}_{\partial U}$ be a subset of the space of boundary data,  let $\mathfrak{S}$ be a functional on $\mathscr{K}_{\partial U}$, and consider the image $\mathscr{M}_{\partial U} := \mathbf{d}\mathfrak{S}( \mathscr{K}_{\partial U} )$.  Then $\mathscr{M}_{\partial U}$ is an isotropic submanifold of $T^\ast \mathscr{Y}_{\partial U}$.  
\end{theorem}

In many cases (for instance, when $\mathcal{Y}_{\partial U}$ is a Hilbert space), it is possible to show that $\mathscr{M}_{\partial U}$ is a Lagrangian subspace of $T^\ast \mathscr{Y}_{\partial U}$ (see \cite{LaSnTu1975, KiTu1979}).  However, for our purposes it will suffice that  $\mathscr{M}_{\partial U}$ is merely isotropic.

By analogy with mechanics, we refer to any functional $\mathfrak{S}$ on the space of boundary data $\mathscr{Y}_{\partial U}$ as a {\bfi Type-I generating functional}.  We now make this analogy more precise.  Consider a mechanical system with configuration space $Q$, and identify the space of boundary data with $Q \times Q$ and the space of normal momenta with $T^\ast Q \times T^\ast Q$, equipped with the symplectic form $- \Omega_0 \oplus \Omega_1$, where $\Omega_i$, $i = 0,1$, is the canonical symplectic form on the $i$th factor.  A function $S(q_0, q_1)$ on $Q \times Q$ then generates a symplectic transformation 
\[
	\left( q_0, \frac{\partial S}{\partial q_0} \right) \mapsto 
		\left( q_1, \frac{\partial S}{\partial q_1} \right)
\]
in the standard sense, and our definition is an extension of this concept to classical field theories.

\paragraph{The Boundary Lagrangian as a Type-I Generating Functional.}  Clearly, the boundary Lagrangian $L_{\partial U}: \mathscr{Y}_{\partial U} \to \mathbb{R}$ is a Type-I generating function in the sense described above.   We let $\mathscr{M}_{\partial U}$ be the image of $\mathbf{d} L_{\partial U}$, restricted to the space of admissible boundary conditions $\mathscr{K}_{\partial U}$.  We can identify $\mathscr{M}_{\partial U}$ with $\mathscr{K}_{\partial U}$, and under this identification the restriction to $\mathscr{M}_{\partial U}$ of the symplectic form $\omega$ on $T^\ast \mathscr{Y}_{\partial U}$ is given by the pull-back form $(\mathbf{d} L_{\partial U})^\ast \omega$.  From Theorem~\ref{thm:isotropy}, we know that $\mathscr{M}_{\partial U}$  is isotropic, so that this form vanishes. However, some interesting results can be obtained by explicitly writing out $(\mathbf{d} L_{\partial U})^\ast \omega$ and equating the result with zero.

The canonical symplectic form $\omega$ is given by $\omega = - \mathbf{d} \Theta$, where $\Theta$ is the canonical one-form on $T^\ast \mathscr{Y}_{\partial U}$ defined in \eqref{canonform}.  The pullback of $\Theta$ along $\mathbf{d} L_{\partial U}: \mathscr{Y}_{\partial U} \to T^\ast \mathscr{Y}_{\partial U}$ then satisfies $(\mathbf{d} L_{\partial U})^\ast \Theta = \mathbf{d} L_{\partial U}$ because of \eqref{pullback_property}, so that 
\[
	\big( (\mathbf{d} L_{\partial U})^\ast \Theta\big)(\varphi) \cdot \delta \varphi = \mathbf{d} L_{\partial U}(\varphi) \cdot \delta \varphi = \int_{\partial U} (j^1 \phi)^\ast \left( \mathbf{i}_{j^1 V} \Theta_L \right),
\]
where we have used \eqref{dl}.  Here, $\phi$ is again the unique solution of the Euler--Lagrange equations with boundary data $\varphi$, and $V$ is a first variation of $\phi$ induced by the boundary variation $\delta \varphi$.  By taking exterior derivatives of both sides and using \eqref{ddl}, we obtain
\begin{equation} \label{iso_to_msff}
	\big( (\mathbf{d} L_{\partial U})^\ast \omega \big)(\varphi) \cdot ( \delta\varphi, \delta\varphi') = 
	\int_{\partial U} (j^1 \phi)^\ast \left( \mathbf{i}_{j^1 W} \mathbf{i}_{j^1 V} \Omega_L \right),
\end{equation}
but by isotropy, we have that the left-hand side of this formula has to vanish. We summarize these results in the following theorem.
\begin{theorem} \label{thm:msff}
	The manifold  $\mathscr{M}_{\partial U} := \mathbf{d} L_{\partial U} ( \mathscr{K}_{\partial U} )$ of admissible boundary data is an isotropic submanifold of $T^\ast \mathscr{Y}_{\partial U}$ if and only if the multisymplectic form formula \eqref{msff} holds.
\end{theorem}

\paragraph{The De Donder--Weyl Equations.}  Given a Lagrangian density $\mathcal{L} = L(x^\mu, y^a, y^a_\mu) \, dV$, we define the multi-momenta $p_a^\mu$ (where $a = 1, \ldots, k$, $\mu = 0, \ldots, n$) and a scalar momentum $p$ as 
\begin{equation} \label{momenta}
	p_a^\mu = \frac{\partial L}{\partial y^a_\mu}
	\quad \text{and} \quad 
	p = L - \frac{\partial L}{\partial y^a_\mu} y^a_\mu.
\end{equation}
These momenta can be defined intrinsically by considering the Legendre transformation as a map from the jet bundle to its extended dual (see, for instance, \cite{CaCrIb1991}).   We now introduce the Hamiltonian function $\mathcal{H}(x^\mu, y^a, p, p_a^\mu)$ as 
\begin{equation} \label{curly_H}
	\mathcal{H}(x^\mu, y^a, p, p_a^\mu) = \underset{y^a_\mu}{\ext} \left[
		p + p_a^\mu y^a_\mu - L(x^\mu, y^a, y^a_\mu) \right],
\end{equation}
where we take the extremum over all values of $y^a_\mu$.  For a related, coordinate-invariant definition of $\mathcal{H}$, we refer to \cite{BiSnFi1988}.  Note that $\mathcal{H}$ vanishes identically on the image of the Legendre transformation \eqref{momenta}, and that $L$ does not necessarily have to hyperregular for $\mathcal{H}$ to be defined.

For the remainder of this paper, we will focus on the locally defined function $H(x^\mu, y^a, p_a^\mu) = p_a^\mu y^a_\mu - L(x^\mu, y^a, y^a_\mu)$, which we will term the {\bfi multi-Hamiltonian (function)}.  We emphasize, however, that unlike $\mathcal{H}$,  $H$ is in general only locally defined, but all the objects defined in this section can be defined in terms of $\mathcal{H}$ solely.

We now rewrite the action density $\mathcal{S}$ in terms of the Hamiltonian function:
\[
	\mathcal{S}_H(y^a, p_a^\mu) = \int_{U} 
		(p_a^\mu y^a_{,\mu} - H(x^\mu, y^a, p_a^\mu) ) \, d^{n+1}x
\]
and we notice that by taking variations of $y^a$ and $p_a^\mu$, we obtain 
\begin{align}
	D \mathcal{S}_H(y^a, p_a^\mu) \cdot (\delta y^a, \delta p_a^\mu) 
		 = &
		\int_U \left(  -\left( \frac{\partial p_a^\mu}{\partial x^\mu} + \frac{\partial H}{\partial y^a} \right) 
		\delta y^a 
		+ 
		\left(  \frac{\partial y^a}{\partial x^\mu} - \frac{\partial H}{\partial p_a^\mu} \right) \delta p_a^\mu \right) \, 
			d^{n+1}x  \nonumber  \\
		& \qquad + \int_{\partial U} p^a_\mu \delta y^a \, d^n x_\mu. \label{DSH}
\end{align}
Under the condition that the variation $\delta y^a$ vanish on the boundary $\partial U$, we obtain the following set of partial differential equations (referred to as the {\bfi De Donder--Weyl equations}):
\begin{equation} \label{DDW}
	\frac{\partial y^a}{\partial x^\mu} = \frac{\partial H}{\partial p_a^\mu}, \quad \text{and} \quad
	\frac{\partial p_a^\mu}{\partial x^\mu} = - \frac{\partial H}{\partial y^a}.
\end{equation}

\paragraph{Type-II Generating Functionals.} We consider again a fixed domain $U \subset X$ and we divide the boundary $\partial U$ into two disjoint parts $A$ and $B$: $\partial U = A \cup B$.   We suppose that, on $A$, we are given fixed boundary fields $\varphi_A$, while on $B$ we are given fixed normal momenta $\pi_B$.  We recall that the components of the normal momenta can be expressed as 
\[
	(\pi_B)_a = p_a^\mu \,  \iota^\ast (d^n x_\mu),
\]
where $\iota : \partial U \hookrightarrow X$ is the inclusion of the boundary in $X$.

For given boundary data $(\varphi_A, \pi_B)$, let $(\phi^a(x), p_a^\mu(x))$ be the solution of the De Donder--Weyl equations \eqref{DDW} with those boundary data, and define the functional 
\begin{align}
	H_{\partial U}(\varphi_A, \pi_B) & =  - \mathcal{S}_H(\phi^a, p_a^\mu)
	+ \int_{B} (\pi_B) \cdot  \phi_{|B} \nonumber \\
	& =-\int_{U}(p_{a}^{\  \mu}\phi_{,\mu}^{a}-H(\phi^{a}, p_{a}^{ \ \mu})) \, d^{n+1}x	
	+  \int_{B}p_{a}^{ \ \mu}\phi^{a} \, d^{n}x_{\mu},\label{boundaryHamiltonian}
\end{align}
which we refer to as the {\bfi boundary Hamiltonian}.
We now compute the derivative of $H_{\partial U}$, keeping in mind the boundary conditions, so that $\delta \varphi_{|A} = \delta \pi_{|B} = 0$.  A similar computation as for the derivation of \eqref{DSH} yields 
\begin{align*}
	DH_{\partial U}(\varphi_A, \pi_B) \cdot (\delta \varphi_A, \delta \pi_B)
	 =& 
		\int_U \left( -\left( \frac{\partial p_a^\mu}{\partial x^\mu} + \frac{\partial H}{\partial y^a} \right) 
		\delta y^a 
		+ 
		\left(  \frac{\partial y^a}{\partial x^\mu} - \frac{\partial H}{\partial p_a^\mu} \right) \delta p_a^\mu \right) \, 
			d^{n+1}x  \nonumber  \\
	&\qquad + \int_B \delta p_a^\mu y^a \, d^n x_\mu - \int_A p_a^\mu \delta y^a \, d^n x_\mu. 
\end{align*}
The integral over the interior vanishes since $(\phi^a, p_a^\mu)$ is a solution of the De Donder--Weyl equations, and the boundary integrals can be written in terms of the normal momenta as 
\[
	D H_{\partial U}(\varphi_A, \pi_B) \cdot (\delta \varphi_A, \delta \pi_B)
	= \int_B \delta \pi_B \cdot \phi_{|B} - \int_A \pi_{|A} \cdot \delta \varphi_A, 
\]
so that the variational derivatives are given by 
\begin{equation} \label{generating_function_2}
	\frac{\delta H_{\partial U}}{\delta \varphi_A} = -\pi_{|A}, 
	\quad \text{and} \quad 
	\frac{\delta H_{\partial U}}{\delta \pi_B} =  \phi_{|B}.
\end{equation}

We compare this with the case of mechanical systems.  Given a finite time interval $[t_0, t_1]$, a Type-II generating function is a function $S(q_0, p_1)$ depending on the position variables $q_0$ at the initial time $t_0$ and on the momenta $p_1$ at the final time $t_1$.  The final position $q_1$ and the initial momentum $p_0$ are then defined by
\[
	 p_0 = \frac{\partial S}{\partial q_0}  \quad \text{and} \quad q_1 = \frac{\partial S}{\partial p_1}.
\]
The relations \eqref{generating_function_2} are the analogue of these expressions for field theory, where the relative minus sign is again due to the orientation of the boundary.

\begin{remark}
In the definition of the boundary Hamiltonian \eqref{boundaryHamiltonian}, we have silently assumed that the De Donder--Weyl equations \eqref{DDW} have a unique solution.  When this is not the case, the boundary Hamiltonian may still be defined: as the Hamiltonian $\mathcal{H}$ in \eqref{curly_H} comes from a Lagrangian $L$, it is sufficient for this that the domain of definition $U$ be sufficiently small; see Remark~\ref{rem:degenerate_lagrangian}.
\end{remark}

\paragraph{The role of the normal momenta.}

One of the difficulties with the covariant De Donder--Weyl equations \eqref{DDW} is the fact that they are severely underdetermined, as there are only $k$ equations for $k(n + 1)$ multimomenta $p_a^\mu$.  One way of addressing this issue is to express the multimomenta $p_a^\mu$ in terms of the derivatives $y^a_\mu$, and to use the fact that mixed partial derivatives are equal, $y^a_{\mu\nu} = y^a_{\nu\mu}$, to arrive at extra differential equations for the multimomenta (referred to as \emph{zero-curvature conditions} in \cite{HoIvPe2012}).

From the definition \eqref{normalmomentum}, we see that the normal momenta $\pi$ --- in contrast to the multimomenta $p_a^\mu$ --- carry a clear physical meaning: the normal momenta $\pi$ are conjugate to the fields $\phi$ on the boundary, and at every point of the boundary, there is one normal momentum for every field.  The pairs $(\phi, \pi)$ of boundary variables therefore represent the true dynamical variables of the theory.  This is also important from a numerical point of view, as describing the evolution of $\phi$ and $\pi$ via \eqref{generating_function_2} is --- at least conceptually --- easier than keeping track of all the multi-momenta $p_a^\mu$ via the De Donder--Weyl equation \eqref{DDW}.

\subsection*{Examples}

\paragraph{Connection to Discrete Variational Mechanics.}
The boundary Lagrangian $L_{\partial U}$ can be viewed as the analogue in Lagrangian field theory of the Jacobi solution of the Hamilton--Jacobi equation. In turn, the Jacobi solution is related to the exact discrete Lagrangian $L_d:Q\times Q\rightarrow\mathbb{R}$ of discrete variational mechanics, which is given by
\begin{equation}
L_d^E(q_0,q_1;h)=\ext_{\substack{q\in C^2([0,h],Q) \\ q(0)=q_0, q(h)=q_1}} \int_0^h L(q(t), \dot q(t)) dt.\label{exact_Ld_variational}
\end{equation}
When the Lagrangian is non-degenerate, this is equivalent to usual characterization of the Jacobi solution,
\begin{equation}
L_d^E(q_0,q_1;h)=\int_0^h L(q_{01}(t),\dot q_{01}(t)) dt,\label{exact_Ld_Jacobi}
\end{equation}
where $q_{01}(0)=q_0,$ $q_{01}(h)=q_1,$ and $q_{01}$ satisfies the
Euler--Lagrange equation in the time interval $(0,h)$. As described in \cite{LeSh2012}, these two characterizations of the exact discrete Lagrangian lead to systematic techniques for constructing computable discrete Lagrangians.

Given a discrete Lagrangian, the discrete Hamilton's principle yields the implicit discrete Euler--Lagrange equations,
\begin{equation}
p_k=-D_1 L_d(q_k, q_{k+1}),\qquad p_{k+1}=D_2 L_d(q_k, q_{k+1}),\label{IDEL}
\end{equation}
which implicitly defines the discrete Hamiltonian map $\tilde{F}_{L_d}:(q_k,p_k)\mapsto(q_{k+1},p_{k+1})$, where the discrete Lagrangian is the Type I generating function of the symplectic transformation.

The exact discrete Lagrangian also greatly simplifies the process of analyzing the order of accuracy of a variational integrator. In particular, Theorem 2.3.1 of \cite{MaWe2001} states that if a discrete Lagrangian, $L_d:Q\times Q\rightarrow\mathbb{R}$, approximates the exact discrete Lagrangian, $L_d^E:Q\times Q\rightarrow\mathbb{R}$ to order $p$, i.e.,
\[ L_d(q_0, q_1;h)=L_d^E(q_0,q_1;h)+\mathcal{O}(h^{p+1}),\]
then the discrete Hamiltonian map, $\tilde{F}_{L_d}:(q_k,p_k)\mapsto(q_{k+1},p_{k+1})$, viewed as a one-step method, is order $p$ accurate.

There is a corresponding notion for Type II generating functions, where we introduce an exact discrete Hamiltonian \cite{LeZh2011},
\begin{equation}
H_d^{E+}(q_k,p_{k+1})=
\ext_{\substack{(q, p) \in
C^2([t_k,t_{k+1}],T^*Q)\\q(t_k)=q_k, p(t_{k+1})=p_{k+1}}} p(t_{k+1}) q (t_{k+1}) - \int_{t_k}^{t_{k+1}} \left[ p \dot{q}-H(q, p) \right] \, dt\label{exact_Hd+}
\end{equation}
where the discrete Hamiltonian map $\tilde{F}_{H_d}:(q_k,p_k)\mapsto(q_{k+1},p_{k+1})$ is defined implicitly by,
\begin{equation}
q_k=D_2 H_d^+(q_{k-1},p_k),\qquad p_k=D_1 H_d^+(q_k, p_{k+1}).\label{IDEL2}
\end{equation}

Clearly, the definition of the boundary Lagrangian $L_{\partial U}$ \eqref{Lext} generalizes the variational characterization of the exact discrete Lagrangian \eqref{exact_Ld_variational}, and the condition that the normal momentum is defined by \eqref{normalmomentum} at every point on the spacetime boundary $\partial U$ generalizes the characterization of a symplectic map in terms of a Type I generating function \eqref{IDEL}. In an analogous fashion, the boundary Hamiltonian \eqref{boundaryHamiltonian} generalizes the exact discrete Hamiltonian \eqref{exact_Hd+}, and the variational derivatives \eqref{generating_function_2} generalize the characterization of a symplectic map in terms of a Type II generating function \eqref{IDEL2}.

Given the important role played by the exact discrete Lagrangian and exact discrete Hamiltonian in the variational error analysis of variational integrators, it is expected that the boundary Lagrangian and boundary Hamiltonian will play a similarly pivotal role in the variational error analysis of discrete Lagrangian and Hamiltonian field theories.

\paragraph{Maxwell's Equations.} 

In this paragraph, we construct the boundary Lagrangian for electromagnetism, and we show that the fact that the image of $\mathbf{d} L$ is isotropic (Theorem~\ref{thm:isotropy}) is equivalent to the statement of {\bfi Lorentz's reciprocity principle} (see \cite{Ji2010}). Note that in this example, the boundary Lagrangian cannot be computed explicitly, but there is in fact no need to do so.

To make the connection with the Lorentz reciprocity principle, we assume that the electric and magnetic fields are simple harmonic functions of time with the same period $\omega$, so that 
\begin{equation} \label{one_ft_mode}
	\mathbf{E}(\mathbf{x}, t) = 
		\hat{\mathbf{E}}(\mathbf{x}) \exp(\mathrm{i} \omega t)
	\quad \text{and} \quad 
	\mathbf{B}(\mathbf{x}, t) = 
		\hat{\mathbf{B}}(\mathbf{x}) \exp(\mathrm{i} \omega t). 
\end{equation}		
		We now work exclusively with the coefficients $\hat{\mathbf{E}}, \hat{\mathbf{B}}$ and we omit the hat over these quantities in the remainder of our treatment.  In terms of these fields, Maxwell's equations in vacuum become 
\[
	\mathrm{i} \omega \epsilon_0 {\mathbf{E}} - \frac{1}{\mu_0} \nabla \times {\mathbf{B}} = 0, 
		\quad \text{and} \quad 
	 \mathrm{i} \omega {\mathbf{B}} + \nabla \times {\mathbf{E}} = 0
\]
together with the constraints $\nabla \cdot {\mathbf{E}} = 0$ and $\nabla \cdot {\mathbf{B}} = 0$. The relation between $\mathbf{E}, \mathbf{B}$ and the vector potential $\mathbf{A}$ can be expressed as 
\begin{equation} \label{gauge_rel}
	 {\mathbf{E}} = - \mathrm{i} \omega  {\mathbf{A}}
		\quad \text{and} \quad 
	 {\mathbf{B}} = \nabla \times  {\mathbf{A}}.
\end{equation}

Maxwell's equations can be derived from the action 
\begin{equation} \label{maxwell_ft_action}
	S({\mathbf{E}}, {\mathbf{B}}) = \int_V \left( 
		\frac{\epsilon_0}{2} {\mathbf{E}}(\mathbf{x})^2 + 
		\frac{1}{2\mu_0} {\mathbf{B}}(\mathbf{x})^2 \right) \, dV
\end{equation}
subject to variations that preserve \eqref{gauge_rel}, viz. $\delta {\mathbf{E}} = - \mathrm{i} \omega \delta {\mathbf{A}}$ and $\delta {\mathbf{B}} = \nabla \times \delta {\mathbf{A}}$.

For any fixed volume $V$ with boundary $S = \partial V$, we denote by $\mathbf{A}_{|S}$ the vector potential restricted to $S$, and likewise for the quantities $\mathbf{E}_{|S}$ and $\mathbf{B}_{|S}$. The boundary Lagrangian is then given by 
\[
	L(\mathbf{A}_{|S}) = \int_V \left( 
		\frac{\epsilon_0}{2} {\mathbf{E}}(\mathbf{x})^2 + 
		\frac{1}{2\mu_0} {\mathbf{B}}(\mathbf{x})^2 \right) \, dV,
\]
where $\mathbf{E}$ and $\mathbf{B}$ are solutions to Maxwell's equations in the interior of $V$ such that \eqref{gauge_rel} holds on the boundary $S$. 

In this example, the space of boundary data $\mathscr{Y}_S$ is the set of all vector potentials $\mathbf{A}_{|S}$ on $S$, and its tangent bundle $T\mathscr{Y}_S$ can be identified with the set of pairs $(\mathbf{A}_{|S}, \delta \mathbf{A}_{|S})$. Using the Euclidian metric, we identify the cotangent bundle $T^\ast \mathscr{Y}_S$ with $T \mathscr{Y}_S$, and we denote its elements by $(\mathbf{A}_{|S}, \mathbf{\Pi}_{|S})$.

By a similar reasoning as the one leading up to \eqref{derformula}, we obtain that the variational derivative of the boundary Lagrangian is given by 
\[
	\frac{\delta L}{\delta \mathbf{A}_{|S}} = 
		\frac{1}{\mu_0} \mathbf{n} \times \mathbf{B}_{|S},
\]
where $\mathbf{n}$ is the unit external normal to $S$, so that the image of $\mathbf{d} L$ is the submanifold 
\[
	\mathrm{Im}(\mathbf{d} L) = 
		\left\{ 
			\left( \mathbf{A}_{|S}, 
			\frac{1}{\mu_0} \mathbf{n} \times \mathbf{B}_{|S}
			\right), \quad \mathbf{A}_{|S} \in \mathscr{Y}_S
		\right\} \subset T^\ast \mathscr{Y}_S.
\]

The symplectic form on $T^\ast \mathscr{Y}_S$ is given by 
\[
	\omega((\delta \mathbf{A}_{|S}, \delta \mathbf{\Pi}_{|S}), 
		(\delta \mathbf{A}'_{|S}, \delta \mathbf{\Pi}'_{|S})) 
	= \int_S \left( \delta \mathbf{A}_{|S} \cdot  \delta \mathbf{\Pi}'_{|S}
		- 
		\delta \mathbf{A}'_{|S} \cdot  \delta \mathbf{\Pi}_{|S}
	\right) dS,
\]
and its pullback along $\mathbf{d} L$ to $\mathscr{Y}_S$ is then
\[
	((\mathbf{d} L)^\ast \omega)(\mathbf{A}_{|S})(\delta \mathbf{A}_{|S}, \delta \mathbf{A}'_{|S}) = \frac{1}{\mu_0} \int_S \left( \delta \mathbf{A}_{|S} \cdot (\mathbf{n} \times \delta \mathbf{B}'_{|S}) -  \delta \mathbf{A}'_{|S} \cdot (\mathbf{n} \times \delta \mathbf{B}_{|S}) \right) \, dS.
\]
The left-hand side is identically zero, so that this equality becomes, using the fact that $\delta \mathbf{E} = - \mathrm{i} \omega \delta \mathbf{A}$, 
\[
	\int_S \left( \delta \mathbf{E}_{|S} \times \delta \mathbf{B}'_{|S} -  
		\delta\mathbf{E}'_{|S} \times \delta\mathbf{B}_{|S} \right)
		 \cdot d\mathbf{S} = 0.
\]
This is precisely the statement of Lorentz reciprocity for vacuum electrodynamics. When sources are present, they can be incorporated in the action \eqref{maxwell_ft_action}, and the full Lorentz reciprocity principle can be derived along the same lines as presented here.

\paragraph{Some further remarks on electromagnetism.}  We finish with a number of remarks on multisymplecticity and Maxwell's equations.

\begin{enumerate}
\item The assumption \eqref{one_ft_mode} was made in order to recover the Lorentz reciprocity principle. When all Fourier modes are present, the multisymplectic form formula is still valid, but more complicated. 

\item 
The construction of the boundary Lagrangian can be extended to the case of discrete Maxwell's equations defined on a space-time mesh $K$ with boundary $\partial K$, as long as the discrete formulation has the same gauge symmetries as the underlying continuous theory, as is the case in the formulation of Bossavit \cite{Bo1998} and the mixed finite element methods based on finite-element exterior calculus of \cite{ArFaWi2010}.  The electromagnetic potential $A$ is then a discrete one-form on $K$ and as a result, the boundary Lagrangian $L_{\partial K}$ is a functional on the space of discrete one-forms on $\partial K$.  As in the next section, this functional can be shown to generate a canonical transformation in the space of boundary data.
 
We also wish to stress that multisymplecticity is a priori not related to the choice of approximation space used to discretize Maxwell's equations: some integration schemes might be multisymplectic (such as Yee's scheme, as shown in \cite{StToDeMa2008}), while others might not be.  The advantage of our formulation is that it gives a conceptually simple criterion to decide upon multisymplecticity: whenever the space of admissible boundary data forms an isotropic submanifold of the space of all boundary data, multisymplecticity is guaranteed.

It is important to note that multisymplectic discretizations and compatible discretizations based on exterior calculus are not mutually exclusive, and there is nothing that precludes the use of approximation spaces based on discrete exterior calculus \cite{Hi2003, Le2004} or finite-element exterior calculus \cite{ArFaWi2010} in the context of a multisymplectic discretization. In particular, discrete exterior calculus was combined with multisymplectic discretizations in \cite{Hi2003,Le2004} to recover Yee's scheme for rectangular meshes, as well as generalizing it to simplicial meshes. Furthermore, the extension to asynchronous time-stepping was developed in \cite{StToDeMa2008}. While the synthesis of multisymplectic discretizations with approximation spaces based on finite-element exterior calculus has not yet been attempted, there is nothing that prevents one from doing so.

\end{enumerate}

\section{Example: Euler Discretization of the Wave Equation}

In this section, we show that concepts such as the boundary Lagrangian can also be defined for discrete field theories.  Under a few modest assumptions on the discretizations used, we recover, among other things, the discrete multisymplectic form formula of \cite{MaPaSh1998}.  Throughout this section, we use the Euler discretization of the linear wave Lagrangian as a motivating example.  For this special case, we discuss the influence of the geometry of the boundary on the expression for the multisymplectic form formula, and we show that the latter reduces to the  multisymplectic conservation law derived by Bridges et al. \cite{Br1997, BrRe2001, BrRe2006}.   Our treatment is inspired by the one in \cite{KoSh2000, MaPaSh1998}, where different discretizations and more complex field theories are treated.  

For the sake of convenience, we restrict ourselves to a scalar field theory with two independent variables, which we label by $t$ and $x$.  We assume that we are given a regular quadrangular mesh in the base space, with mesh lengths $\Delta t$ and $\Delta x$, and we denote the nodes in this mesh by $(n, i) \in \mathbb{Z} \times \mathbb{Z}$.  Note that the node $(n, i)$ corresponds to the point $x^n_i := (n\Delta t, i \Delta x)$ in $\mathbb{R}^2$.  We denote the value of the field $u$ at the node $(n, i)$ by $u^n_i$.  

As in \cite{MaPaSh1998}, we introduce a discrete version of the jet bundle as follows.  We define a {\bfi triangle} at $(n, i)$ to be the ordered triple $((n, i), (n, i+1), (n+1, i))$, which we denote by $\bigtriangleup^n_i$ (or simply by $\bigtriangleup$ if no confusion can arise), and we let $X_\bigtriangleup$ be the set of all such triangles.  Given a triangle $\bigtriangleup = ((n, i), (n, i+1), (n+1, i))$, we refer to the vertices in a concise way by $\bigtriangleup_1 := (n, i)$, $\bigtriangleup_2 := (n, i+1)$, and $\bigtriangleup_3 := (n+1, i)$.

The discrete jet bundle is then given by (see \cite{MaPaSh1998}):
\[
	J^1_\bigtriangleup Y := \{ (u^n_i, u^n_{i+1}, u^{n+1}_i) \in \mathbb{R}^3: 
		((n, i), (n, i+1), (n+1, i)) \in X_\bigtriangleup \},
\]
and so is equal to $X_\bigtriangleup \times \mathbb{R}^3$.  Given an element $(u^n_i, u^n_{i+1}, u^{n+1}_{i})$ of the discrete jet bundle, we define the triangle midpoint by $\bar{u}^n_i := (u^n_i + u^n_{i+1} + u^{n+1}_i)/3$, and we introduce the following first-order expressions for the temporal velocity $v^n_i$ and the spatial velocity $w^n_i$:
\begin{equation} \label{forward_discretizations}
	v^n_i := \frac{u^{n+1}_i - u^n_i}{\Delta t} \quad \text{and} \quad
	 w^n_i := \frac{u^n_{i+1} - u^n_i}{\Delta x}.
\end{equation}
Consequently, we can discretize a given Lagrangian density $\mathcal{L}$ on $J^1 Y$ by  
\begin{equation}\label{discreteLtri}
L_{d}(u^n_i, u^n_{i+1}, u^{n+1}_{i})
:= \frac{\Delta t\Delta x}{2}
L\left(\frac{u^{n+1}_i - u^n_i}{\Delta t}, \frac{u^n_{i+1} - u^n_i}{\Delta x}, \frac{u^n_i + u^n_{i+1} + u^{n+1}_i}{3}\right).
\end{equation}
For the linear wave equation, with Lagrangian \eqref{wave_lagrangian}, the discrete Lagrangian becomes
\[
	L_{d}(u^n_i, u^n_{i+1}, u^{n+1}_{i}) = 
		\frac{\Delta t \Delta x}{4} \left( 
			\left(\frac{u^{n+1}_i - u^n_i}{\Delta t}\right)^2
			- \left(\frac{u^n_{i+1} - u^n_i}{\Delta x}\right)^2
		\right).
\]

\paragraph{Poincar\'e--Cartan Forms.}  Given a discrete Lagrangian $L_d$, we introduce the discrete Poincar\'e--Cartan forms $\Theta^{1}_{L_d}$, $\Theta^2_{L_d}$ and $\Theta^3_{L_d}$ by 
\[
	\Theta^{1}_{L_d} (u^n_i, u^n_{i+1}, u^{n+1}_{i}) := D_1 L_d (u^n_i, u^n_{i+1}, u^{n+1}_{i}) \, d u^n_i,
\]
and similarly for $\Theta^2_{L_d}$ and $\Theta^3_{L_d}$.  Note that these forms are one-forms on the discrete jet bundle, and that $\Theta^1_{L_d} + \Theta^2_{L_d} + \Theta^3_{L_d} = \mathbf{d} L_d$.  Furthermore, we put $\Omega^k_{L_d} = - \mathbf{d} \Theta^k_{L_d}$ (for $k = 1, 2, 3$), so that 
\begin{equation}
	\Omega^1_{L_d} + \Omega^2_{L_d} + \Omega^3_{L_d} = 0.
\end{equation}

For the linear wave equation, a straightforward computation yields
\begin{equation} \label{Omega1}
	\Omega^1_{L_d} = 
	\frac{1}{2} ( (\Delta x) \, dv^n_i \wedge du^n_i - (\Delta t) \, dw^n_i \wedge du^n_i) 
\end{equation}
as well as  
\begin{equation} \label{Omega23}
	\Omega^2_{L_d} = \frac{1}{2} (\Delta t) \, dw^n_i \wedge du^n_i,
	\quad \text{and} \quad 
	\Omega^3_{L_d} = - \frac{1}{2} (\Delta x) \, dv^n_i \wedge du^n_i,
\end{equation}
where $v^n_i$ and $w^n_i$ are given by \eqref{forward_discretizations}.

\paragraph{Discrete Euler--Lagrange Equations.}

Given a finite subset $U \subset X_\bigtriangleup$ of the space of triangles, we form the {\bfi discrete action sum} as
\[
	\mathcal{S}_U(u) := \sum_{\bigtriangleup \in U} L_d( j^1 u(\bigtriangleup)).
\]
Here, $u$ is a {\bfi discrete field}, assigning to every node $(n, i)$ a field value $u^n_i$, and $j^1 u$ is its {\bfi first jet extension}, defined by 
\[
	(j^1 u)(\bigtriangleup) = (u(\bigtriangleup_1), u(\bigtriangleup_2), u(\bigtriangleup_3)) \in J^1_\bigtriangleup Y.
\]

\begin{figure}
	\begin{center}
		\includegraphics{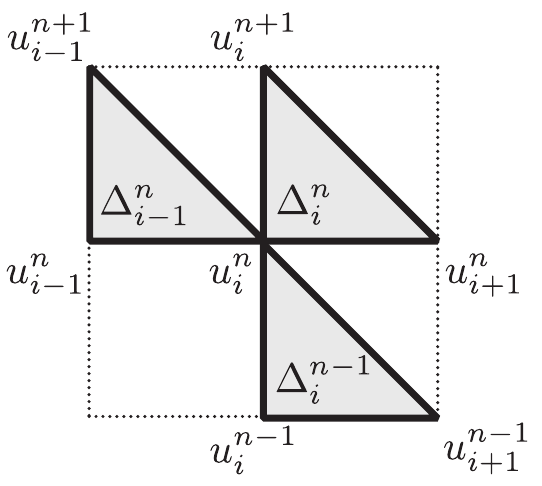}
		\caption{Three adjacent triangles touching a common vertex, labeled by $u^n_i$. \label{fig:triangles}}
	\end{center}
\end{figure}

We now focus on a particular configuration $U$, consisting of three adjacent triangles, as in Figure~\ref{fig:triangles}.  The action sum for this $U$ is explicitly given by 
\begin{equation} \label{discrete_S}
\mathcal{S}_U(u) =
	L_{d}(u^n_i, u^n_{i+1}, u^{n+1}_i)
	+L_{d}(u^n_{i-1}, u^n_i, u^{n+1}_{i-1})
          +L_{d}(u^{n-1}_i, u^{n-1}_{i+1}, u^n_i).
\end{equation}
By keeping the values of the field on the boundary fixed, and taking variations with respect to $u_{n,i}$, we obtain the following {\bfi discrete Euler--Lagrange equations}:
\begin{equation} \label{DEL}
	D_{1}L_{d}(u^n_i, u^n_{i+1}, u^{n+1}_i)
	+D_{2}L_{d}(u^n_{i-1}, u^n_i, u^{n+1}_{i-1})
	+D_{3}L_{d}(u^{n-1}_i, u^{n-1}_{i+1}, u^n_i) = 0.
\end{equation}

We can rewrite these equations in terms of the discrete Poincar\'e--Cartan forms as 
\begin{equation} \label{PC_DEL}
	\Theta^1_{L_d}(\bigtriangleup^n_i) + \Theta^2_{L_d}(\bigtriangleup^n_{i-1}) + \Theta^3_{L_d}(\bigtriangleup^{n-1}_i) = 0, 
\end{equation}
where $\bigtriangleup^n_i$, etc., refer to the triangles defined in Figure~\ref{fig:triangles}.

For the wave equation \eqref{wave_lagrangian}, the discrete Euler--Lagrange equations result in the standard second-order scheme 
\begin{equation} \label{wave_eq_DEL}
-\frac{u_{i}^{n+1}-2u_{i}^{n}+u_{i}^{n-1}}{(\Delta t)^{2}}
+ \frac{u_{i+1}^{n}-2u_{i}^{n}+u_{i-1}^{n}}{(\Delta x)^{2}}
=0.
\end{equation}

\paragraph{Discrete Boundary Lagrangian.}  We now mimic the construction in Section~\ref{sec:boundary_lagrangian} to introduce a discrete version of the boundary Lagrangian.  We refer again to Figure~\ref{fig:triangles} and we define the boundary Lagrangian to be the extremal value of the action sum \eqref{discrete_S}, taking variations over the interior node $u^n_i$.  For the sake of notation, we denote the values of the field on the boundary by $u_{\partial U}$, so that 
\[
	u_{\partial U} := (u^n_{i+1}, u^{n+1}_i, u^{n+1}_{i-1}, 
		u^n_{i-1}, u^{n-1}_i, u^{n-1}_{i+1}).
\]
The boundary Lagrangian is then given by
\begin{equation} \label{discrete_boundary_lagrangian}
	L_{\partial U} (u_{\partial U}) 
:= \underset{{u^n_i}}{\ext} \big[ 
	L_{d}(u^n_i, u^{n}_{i+1}, u^{n+1}_i)
+L_{d}(u^n_{i-1}, u^n_i, u^{n+1}_{i-1})
          +L_{d}(u^{n-1}_i, u^{n-1}_{i+1}, u^n_i)\big],
\end{equation}
where the extremum is taken over all $u^n_i$.   Alternatively, $L_{\partial U}$ can be defined by solving the discrete Euler--Lagrange equations \eqref{DEL} for $u^n_i$ given the values of $u_{\partial U}$ as boundary data, and substituting this value into the action sum \eqref{discrete_S}.  

\paragraph{Discrete Multisymplectic Form Formula.}   

We now derive the discrete multisymplectic form formula by twice taking the exterior derivative of the boundary Lagrangian $L_{\partial U}$.   By taking the exterior derivative of both sides of \eqref{discrete_boundary_lagrangian} and using the definition of the Poincar\'e--Cartan forms given above, we  obtain 
\[
	\mathbf{d} L_{\partial U} = \sum_{k = 1}^3  \sum_{l = 1}^3 \Theta^k_{L_d}( \Delta^{(l)} ).
\]
Using the discrete Euler--Lagrange equations \eqref{PC_DEL}, we can rewrite this as 
\begin{equation} \label{discrete_dl}
	\mathbf{d} L_{\partial U} = \sum_{k = 1}^3  
		\sum_{l=1; l \ne k}^3\Theta^k_{L_d}( \Delta^{(l)} ),
\end{equation}
and by taking another exterior derivative of both sides, and using that $\mathbf{d}^2 L_{\partial U} \equiv 0$, we finally obtain 
\begin{equation} \label{discrete_MSFF}
	0 = \sum_{k = 1}^3  \sum_{l=1; l \ne k}^3 \Omega^k_{L_d}( \Delta^{(l)} ).
\end{equation}
This is precisely the multisymplectic form formula derived in \cite{MaPaSh1998}, applied to the triangular domain of Figure~\ref{fig:triangles}.  For the linear wave equation, we can substitute the expressions \eqref{Omega1} and \eqref{Omega23} for the discrete multisymplectic forms to obtain 
\[
	\frac{1}{\Delta x} (dw^n_i \wedge du^n_i - dw^n_{i-1} \wedge du^n_{i-1})
	- 
	\frac{1}{\Delta t} (dv^n_i \wedge du^n_i - dv^{n-1}_i \wedge du^{n-1}_i) = 0.
\]
This is the multisymplectic conservation law of Bridges and Reich \cite{BrRe2001} applied to the linear wave equation, but we caution against taking this result too far: in general the discrete Euler--Lagrange equations \eqref{DEL} will be different from the discrete multisymplectic equations of Bridges and Reich.  Consequently, the discrete multisymplectic form formula \eqref{discrete_MSFF} will also lead to different multisymplectic conservation laws.  

\paragraph{Lagrangian Submanifolds and Characteristics.}
With the notations of the previous paragraph, let $\mathscr{M}_{\partial U}$ be the image of $\mathbf{d} L_{\partial U}$ given by \eqref{discrete_dl}.  Note that $\mathscr{M}_{\partial U}$ is a submanifold of $(T^\ast Q)^{\times 6}$, where we have one copy of $T^\ast Q$ for each boundary point in Figure~\ref{fig:triangles}.   Note that $\mathscr{M}_{\partial U}$ is the image of an exact one-form, and hence determines at least an isotropic submanifold of $(T^\ast Q)^{\times 6}$.  If there are no extra conditions on the boundary data, then $\mathscr{M}_{\partial U}$ will be Lagrangian.

We now discuss the relation between the location of the boundary data and the nature of $\mathscr{M}_{\partial U}$.  To keep the computations to a minimum, we restrict ourselves to the linear wave equation.  We first solve the discrete wave equations \eqref{wave_eq_DEL} to determine $u^n_i$ in terms of the boundary data.  The discrete wave equation can be rewritten as 
\begin{equation} \label{boundary_rel}
	2(c^2 - 1) u^n_i = c^2(u^n_{i+1} + u^n_{i-1}) - (u^{n+1}_i + u^{n-1}_i)
\end{equation}
where $c := \Delta t/\Delta x$ is the aspect ratio of the mesh.  If the CFL condition is satisfied, so that $c < 1$, this expression can be used to determine $u^n_i$ in terms of the boundary data.  However, when $c = 1$, the left-hand side vanishes and we merely obtain a relation between the boundary data.   In this case, the boundary data are located on characteristics of the continuous wave equation.

As a result, whenever the boundary data is characteristic, there exist supplementary compatibility conditions between the boundary data, and hence $\mathscr{M}_{\partial U}$ is strictly isotropic.  When the boundary is noncharacteristic, there are no further conditions on the data and $\mathscr{M}$ is Lagrangian.  We outlined a similar result for the continuous wave equation at the end of Section~\ref{sec:boundary_lagrangian}.

\section{Conclusion and Future Directions}

In this paper, we consider the space of boundary data for a Lagrangian field theory, whose tangent space is the space of Cauchy data, and we introduce a duality pairing between the space of Cauchy data and normal momenta. We then introduce the boundary Lagrangian, which is the analogue of Jacobi's solution of the Hamilton--Jacobi equation, and consider more generally the concept of Type-I and Type-II generating functionals for a Lagrangian field theory.

We are interested in the following topics for future work:

\begin{itemize}
\item \textit{Continuous and Discrete Hamilton--Jacobi Theory for Field Theories.} By computing variations of Jacobi's solution of the Hamilton--Jacobi equation, one can recover the Hamilton--Jacobi equation. In a similar way, we intend to use the concept of the boundary Lagrangian to systematically derive a Hamilton--Jacobi equation for covariant field theories.

In \cite{OhBlLe2011}, a discrete Jacobi's solution was used to derive a discrete Hamilton--Jacobi equation, which is in turn related to the Hamilton--Jacobi--Bellman equation of optimal control. It would be interesting to develop the analogous connection between discrete Hamilton--Jacobi theory for fields and numerical methods for the optimal boundary control of Lagrangian field theories.

\item \textit{Variational Error Analysis of Variational Integrators for Field Theories.} Since the boundary Lagrangian $L_{\partial U}$ and the boundary Hamiltonian $H_{\partial U}$ are exact generating functionals for Lagrangian and Hamiltonian field theories, it would be natural to extend the theory of variational error analysis introduced in \cite{MaWe2001} to the setting of discrete field theories, and to develop general techniques for constructing variational integrators for field theories by extending the approaches for constructing Lagrangian and Hamiltonian variational integrators described in \cite{LeZh2011, LeSh2012}.

\item \textit{Connections with Multisymplectic Integrators for Hamiltonian PDEs.} A variational characterization of discrete multisymplectic variational integrators for Hamiltonian field theories is a natural direction to pursue, and will involve a combination of the insights developed in this paper, and a generalization of the methods developed in \cite{MaPaSh1998, VaCa2007, LeZh2011}. The goal would be to provide a systematic characterization of the associated discrete multisymplectic conservation laws, and appropriate invariance properties of the discrete Hamiltonian which would lead to a discrete multisymplectic Noether's theorem. This connections between such a discrete variational approach and the multisymplectic integrators described in \cite{BrRe2001, BrRe2006} remain to be elucidated.

\end{itemize}

\paragraph{Acknowledgments.}  We thank M. Gotay and J. \'Sniatycki for useful discussions and for bringing reference \cite{LaSnTu1975} to our attention.    

The authors are partially supported by NSF CAREER award DMS-1010687 and NSF FRG grant DMS-1065972.  C. L. is also supported by the China Scholarship Council. J. V. is on leave from a Postdoctoral Fellowship of the Research Foundation--Flanders (FWO-Vlaanderen). This work is part of the Irses project GEOMECH (nr. 246981) within the 7th European Community Framework Programme.  

\def\polhk#1{\setbox0=\hbox{#1}{\ooalign{\hidewidth
  \lower1.5ex\hbox{`}\hidewidth\crcr\unhbox0}}}


\end{document}